# Cross Waveguide Design for Color-Centers in Diamond for Photonic Quantum Computing

ALESSIO MIRANDA,[1,2,*] RYOICHI ISHIHARA,[1,2] AND SALAHUDDIN NUR[1]

[1]*Department of Quantum Computer Engineering, Faculty of Electrical Engineering Mathematics and Computer Science, Delft University of Technology, Delft, The Netherlands*
[2]*QuTech, Delft University of Technology, Delft, The Netherlands*
*\*a.m.d.miranda@tudelft.nl*

**Abstract:** Color centers in diamond are a promising platform for quantum computing applications because of their optical and spin properties. However, diamond presents some technological challenges that limit its use in complex or large photonic circuits. To mitigate these limitations, it is technically effective to separate the smallest possible diamond photonic structures or chiplet containing the color center(s) from the rest of the circuit, which is fabricated on another material platform, and then heterogeneously integrate them. Considering efficient excitation and photon collection from waveguide-coupled color centers, we design a cross waveguide as the primary component of our chiplet to access the color centers, channeling excitation and emitted photons into different waveguides, and connecting the structure to the other components of the photonic circuit. The chiplet containing the cross waveguide and supporting structures requires careful optimization of each subcomponent. The receptor's design is also critical for optimal signal transmission. In this paper, we develop a simple but efficient methodology to optimize the main components constituting both the chiplet and the receptor for their synergistic operation. The designed structure has an excitation-to-emission conversion of more than 5.4%, crosstalk of less than -40 dB, a working bandwidth of 160 nm, fabrication feasibility and tolerance within the limits of modern nanofabrication, and a mechanical solid structure with a footprint of less than 2000 µm².

## Introduction

Diamond is an exceptional material for power [1] and extreme-environment [2,3] electronics due to its superior electronic [4], thermal [5], mechanical [6], and optical properties [7], including high electrical background field strength [8], carrier mobility [4], thermal conductivity [9], Young's modulus [10,11], and a high refractive index [7,12]. Recently, it has also received attention as a promising material for quantum technologies as a host material for color centers [13–17].

Color centers (CCs) in diamond are particularly attractive for quantum information science because of their long spin coherence times [18], optical readout capabilities [19], and high operational temperatures [20,21]. They are also used for high-precision sensing for measuring magnetic fields [22], temperature [23,24], and pressure [25]. In particular, tin vacancies (SnV), because of their symmetrical structure [26,27,13], have reduced sensitivity to electric field fluctuations and noise [28], good optical properties [29–32] and have good compatibility with nanofabricated photonic structures [33,34]. In addition to this, they show a higher quantum efficiency (80%) than other III/V based color centers [35], and long spin coherence times at temperatures above 1K [30], thus resulting in an excellent candidate for large-scale quantum communication and computing [13,27,36].

However, diamond also presents some challenges in scaling potential quantum systems due to the current lack of possible wafer-scale substrate growth, photonic-grade diamond thin film on low-index oxide, and patterning difficulties arising from the limited size of the substrates, potential incompatibility with materials used in other platforms, and the immaturity of the nanofabrication technology [14,37,38]. This hinders the use of diamond for large and complex photonic circuits for on-chip quantum systems [39–42], which would require a relatively large diamond-on-insulator type substrate and precise nanofabrication.

To overcome the technological limitations of fabrication and monolithic integration of large diamond photonic circuits, it is convenient to separate the small diamond chiplet containing the CCs from the rest of the photonic circuit (fabricated on another platform such as SiN) and then heterogeneously integrate the two (e.g. by pick and place, P&P) [43–46]. The diamond chiplet consists of a multimode interferometric (MMI) cross waveguide used to access the CC and collect its emission and tethers to a supporting frame. The receptor on the other material (SiN) platform consists of adiabatic couplers to those of the chiplet and the remaining photonic components. In addition to P&P, this design is suitable for integration schemes such as transfer printing and DOI monolithic [43]. The cross-shaped structure enables separation of the excitation field from the CC emission, routing them into different waveguides. This facilitates optimization of the device design and operation, and supports its use in quantum internet applications [47,48].

Each component of the chiplet and the receptor must be designed in synergy with the others, taking into account excitation-to-emission conversion, transmission efficiency, operating bandwidth, fabrication feasibility, and tolerance. Here, we propose a methodology for the design and optimization of all key elements of a diamond chiplet for CC-based modular quantum systems, including the receptor and their combined integration. We first determine the diamond waveguide dimensions that support single- or double-TE-mode propagation. We then optimize the cross-shaped multimode interferometer and its tapers to reduce reflections and enhance excitation of the CC located at its center. Next, we investigate the supporting tethers, optimizing their size and position to minimize loss and suppress unwanted oscillations. We also introduce additional structures to facilitate alignment between the diamond chiplet and the SiN receptor. Since this alignment is challenging, we study the transmission limits in detail, with particular focus on the misalignment tolerance of the adiabatic couplers connecting the two parts. Finally, we perform a mechanical analysis to assess whether the tethers can reliably support the weight of the full chiplet, and summarize the results to provide an overview of the optimized devices' properties. Our optimization methodology minimizes degrees of freedom and simplifies fabrication while maintaining strong performance. We also discuss alternative approaches reported in the literature and justify the choices made here. Beyond quantum computing, this methodology is relevant to a broader photonics community interested in optimizing widely used photonic components.

**Structure for a pick and place technology**

The pick and place technology, shown schematically in Fig. 1, allows heterogeneous integration of two photonic components fabricated on separate platforms [45,49,50,46]. The former component, fabricated in a less scalable platform (here, a diamond with CC qubits), is a suspended chiplet attached to a supporting frame via tethers. The latter, called a receptor, is the larger/scalable photonic circuit on which the chiplet will be positioned, and is usually fabricated in a more common platform, such as SiN on $SiO_2$, as in this case. The process of picking consists of gently pressing a needle against the chiplet to break the tethers at specific points (e.g., where they have lower mechanical strength, such as where their width is narrower) and then releasing the chiplet. The chiplet remains attached to the needle because of van der Waals forces.

Subsequently, the chiplet is transferred with the same needle and placed on the receptor; the van der Waals forces between the chiplet and the receptor are stronger than those between the chiplet and the needle because of the larger contact surface, hence the chiplet remains attached to the receptor when positioned hereon.

With reference to Fig. 1 the chiplet, entirely fabricated in diamond, and the receptor, entirely fabricated in SiN on $SiO_2$, are complex structures made of different parts,whose specific role will be clarified in this manuscript and which need to be optimized for a perfect light transmission: the receptor, shown in Fig. 1 (a), consists of adiabatic couplers to be optimized in synergy with the corresponding adiabatic couplers of the chiplet, which couple the light from the rest of the photonic circuit to the chiplet, and alignment marks; the chiplet, shown in Fig. 1b, consists of a frame that supports cross waveguides via tethers. The cross waveguide is the most important part of the design because it embeds the CC and provides access to it for both excitation and emission. The waveguides departing from the cross terminate with adiabatic couplers which matches those of the SiN receptor, the tethers support also alignment marks matching those in the SiN for a visually facilitated alignment, in the center of the cross waveguide there is a multimode interferometer (MMI) optimized to focus power onto the CC while minimizing crosstalk, in this way the excitation (be it a single photon or a continuous wave laser) is sent from the one of the horizontal waveguides, excites the CC via the MMI without propagating through the vertical waveguides, which are used for collecting the emitted light from the CC.

During P&P the needle breaks the tethers along the red dashed line, separating it from the frame, and then, as shown in Fig.1c transfer the pending chiplet to the receptor, it is crucial that each part of the chiplet and receptor is optimized properly to guarantee an efficient alignment and coupling between the two parts, as well as an optimal excitation of the CC and a clean collection of its emission, in the following paragraphs we will discuss this in detail the optimization of each component.

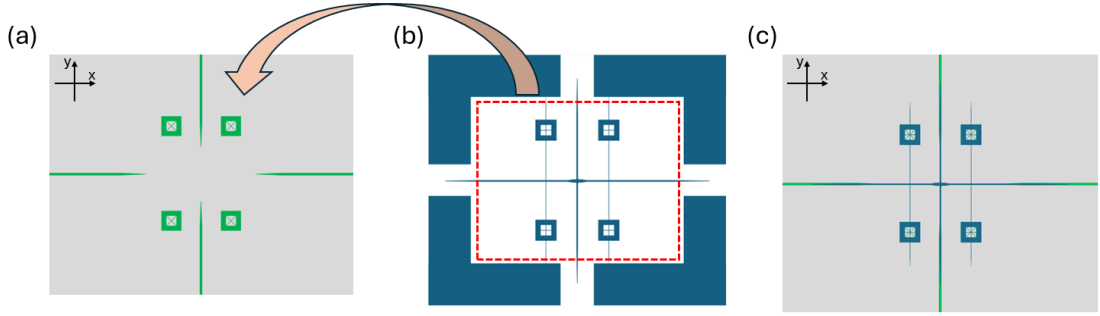

Fig. 1. the operation of pick and place consists in fabricating (a) SiN receptor on SiO$_2$ and (b) a separate diamond chiplet attached to its supporting frame; during operation the chiplet is removed from the frame by cracking the tethers along the dashed square and (c) is positioned on the receptor.

**Methodology of the simulation and optimization of the waveguides**

Simulations are performed using the ANSYS Lumerical [51] software packages for 3D finite-difference time-domain solver (Lumerical-FDTD) and EigenMode Expansion (Lumerical MODE-EME) solver, with a 15nm mesh in all directions. For this study, we chose to operate at the wavelength corresponding to the resonant TE excitation of SnV color centers (CCs), ($\lambda = 620 nm$ [13,35]. This wavelength corresponds to the typical room-temperature emission of SnV CCs in the absence of stress or other environmental perturbations [32]. For simplicity, and to maintain consistency between the two solvers, the emission is assumed to be point-like and spectrally infinitesimal ($\Delta\lambda < 10^{-10} nm$), although the typical full width at half maximum of SnV emission is $\Delta\lambda = 6.2 nm$. As shown below, this approximation does not affect the performance analysis. The results, therefore, remain valid for excitation originating either from a laser source or from another CC.

Due to the limited wavelength range analyzed in this study, we consider all the refractive indices to be constant and equal to those measured at λ = 620 nm: diamond [12] $n_c = 2.4114$, Si$_3$N$_4$ [52,53], $n_{Si_3N_4} = 2.0121$, SiO$_2$ [55,56], $n_{SiO_2} = 1.4574$.

The ports and monitors have a width of 1μm × 1μm for the diamond waveguides and are at least 600 nm wider than the width of the SiN waveguides. The transmissions are calculated by the ratio of the power recorded at the 2μm × 2μm monitors placed perpendicularly to the corresponding direction of propagation of waveguides, while the emission of the CCs, which happens in all directions, is quantified by summing the power recorded of six 50 nm × 50 nm plane monitors disposed as a cube around it.

With reference to Fig. 2, we select diamond waveguides with a width of w = 300 nm, which is common for suspended diamond waveguides [14,29,57], and thicknesses of h = 150 and h = 250 nm. These geometries support only one and two TE-polarized modes, respectively, as shown in Fig. 2 (a). It is also worth noting that, for widths larger than 300 nm, the difference between the effective indices of the 250 nm and 15 nm thick waveguides remains nearly constant at approximately 0.2, as clearly visible in Fig.2 (b). For the SiN waveguides of the receptor, we similarly choose a cross-section of 500 nm × 250 nm, which supports only the fundamental TE mode and is also widely adopted in the literature [58,59].

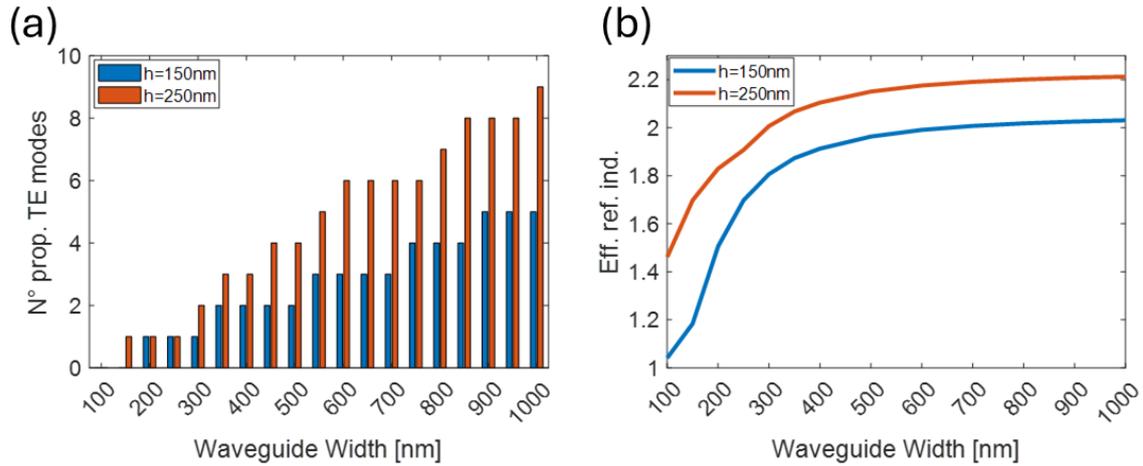

Fig. 2. (a) Dependence of effective refractive index for TE0 mode in diamond waveguides on width for thickness of 150 nm and 250 nm; (b) width for which only the fundamental mode is supported.

## 1. Optimization of the Multimode interferometer in the Cross waveguide

### 1.1 The necessity of focusing on the CC and minimizing crosstalk

Cross waveguides are commonly used in photonics to separate input and output signals because of their relative ease of fabrication, the possibility of integrating photonic crystals into their structure, and their suitability for a resonant excitation scheme. They can also perform advanced operations such as polarization and mode multiplexing or subwavelength operation [60–62].

Here, the CC is located at the center of a cross waveguide, viz., of the chiplet itself, as in Fig. 3 (a). In the zoom of the central part of the cross waveguide, shown in Fig. 3 (b), it can be observed how the structure is used to provide a separate input and output for the excitation (e.g., via the horizontal waveguides) and for the emission (e.g., via the vertical waveguides) of the CC.

To ensure efficient excitation of the CC, most of the power of the incoming excitation light should be concentrated on the CC itself, that is, at the very center of the cross waveguide. At the same time, the crosstalk between orthogonal waveguides used for excitation and emission, respectively, should be minimized.

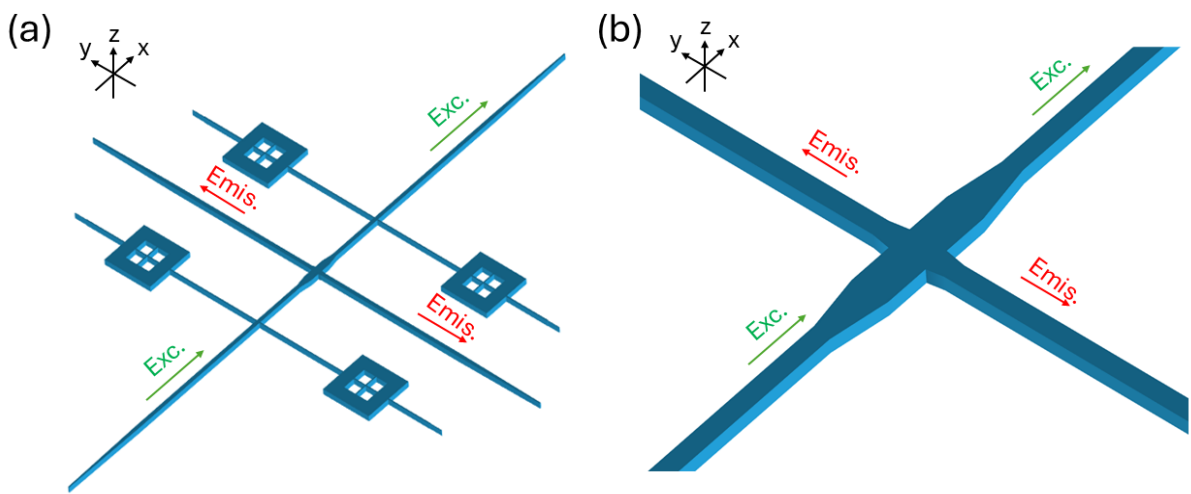

Fig. 3. (a) Bird's eye view of the chiplet with direction of excited and emitted light, the scale is 1:1:2, meaning that the x and y components are equally scaled while the thickness (z) is twice for better visibility; (b) zoom of the center of the chiplet at which the very center of the CC is located.

We achieve these requirements by positioning a multimode interferometer (MMI) at the center of the cross. An MMI works on the basis of self-imaging based on constructive interference of propagating modes [63]. When light is launched into the MMI symmetrically with respect to the sides of the MMI, as shown in Fig. 3 (b), the excitation couples only to the even (symmetric) modes, while the odd (antisymmetric) modes are ideally not excited. In order to obtain constructive interference, it is therefore important that at least the fundamental and the next even mode are guided in the MMI.

The MMI dimensions are selected so that only these two modes dominate, enabling favorable self-imaging while keeping the device footprint small. The relative phase evolution between these two modes is determined by their propagation constants, which define the characteristic beating length as $L_\pi = \frac{\pi}{\beta_0 - \beta_2}$ where $\beta_0$ and $\beta_2$ are their corresponding propagation constants. These can be determined numerically or approximated analytically [63] as $\beta_\nu \simeq \frac{2\pi n_{eff}}{\lambda} - \frac{(\nu+1)^2 \pi \lambda}{4 n_{eff} W^2}$ where $\nu$ is the mode order, $n_{eff}$ is the corresponding effective index, W is the effective propagation width including the lateral penetration depth due to the Goos-Hänchen shift, and $\lambda$ is the vacuum wavelength of the propagating light. Since the physical width of the MMI is moderately large, and with reference to Fig. 2 (b), we further simplify the expression by assuming similar effective refractive indices for the two modes and by taking the effective width to be approximately equal to the physical width of the MMI. The interference between the two modes repeats at lengths [63] $L_p = 3p L_\pi$. The first image is thus formed at $L_1 = 3 L_\pi$. At this position, the two even modes interfere constructively and reproduce the input image, such that the optical field is concentrated at the vertical (y) middle of the MMI, while the power at the sides remains negligible. An MMI of length $L = 2 L_1 = 6 L_\pi$, with a CC positioned at its midpoint, that is, at the center of the reproduced input image, can therefore excite the CC with maximum power. At the same time, the optical power at the corresponding sides remains minimal, thereby reducing the associated crosstalk.

The choice of the MMI width is critical not only for determining the image positions, but also for ensuring adequate image quality. For our application, this quality can be assessed using three criteria: (i) the size of the central image, which should overlap as much as possible with the region defined by the cross section of the CC (estimated to be $10^{-12}$ cm² as discussed below) while still enabling efficient excitation in the presence of possible CC misalignment from its nominal center; (ii) the crosstalk, defined as the fraction of power coupled to the side where the emission waveguide branches off; and (iii) the overall MMI footprint, which should be kept as small as possible. Fig. 4 shows the image distributions for rectangular MMIs with different widths and a length of 4 μm, assuming the input waveguide is located at (0, 0). Figs. 4 (a)-(f) present the results for widths of 500 nm, 600 nm, and 700 nm, each at thicknesses of 150 nm and 250 nm. We begin the analysis at 500 nm because the minimum width required to support the first two even TE modes is approximately 450 nm for both thicknesses. From a qualitative assessment, the optimal width is 600 nm: for smaller widths, the crosstalk is higher, whereas for larger widths, the image separation increases without a significant improvement in image quality.

To obtain a more systematic evaluation, we plot in Fig. 4 (g) and Fig. 4 (i) the power recorded within a 40 nm × 40 nm rectangle positioned at the mid-plane of the MMI and scanned along the x-axis. The peak power in the first image is lower for the 500 nm-wide MMI than for the other widths considered. In Fig. 4h and Fig. 4j, we plot the power recorded along a 320 nm line at the side of the MMI, also positioned at the mid-plane and scanned along the x-axis. This power contributes to crosstalk into the emission waveguides and should therefore be minimized. Again, the 500 nm-wide MMI performs worse than the other two widths. Based on this analysis, we conclude that a 600 nm-wide MMI provides the best compromise among image quality, crosstalk suppression, and footprint, and we therefore select this width for both MMI thicknesses.

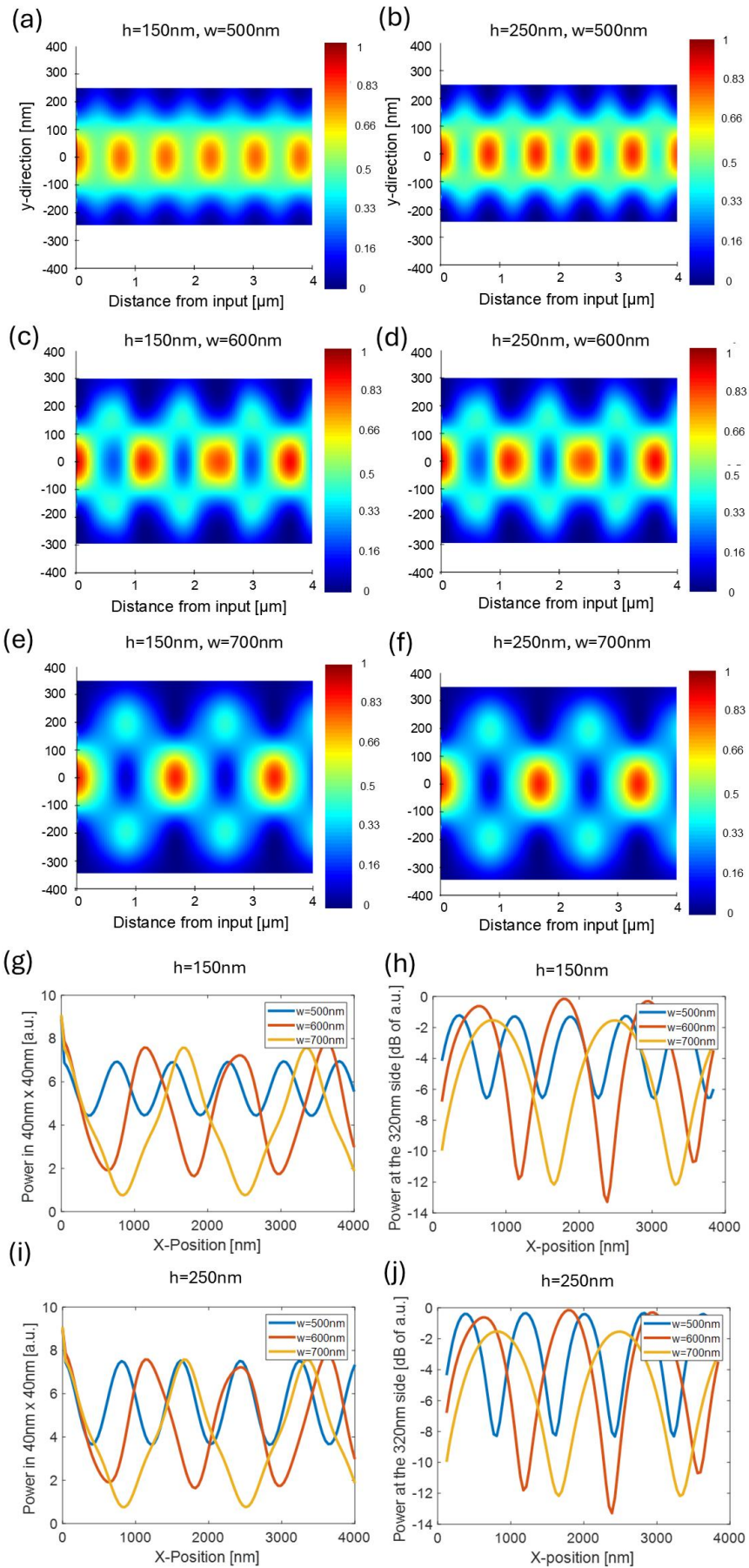

Fig.4. Power distribution of MMIs with different widths (with incoming waveguide of 300 nm width), for the widths of (a), (b) 500 nm, (c), (d) 600 nm, (e), (f) 700 nm for the examined thicknesses of 150 nm and 250 nm respectively. Power distribution for (g) an area 40 nm x 40 nm and (h) for the crosstalk along 320 nm line at the border of the MMI as a function of the x-position for an MMI with thickness 150 nm and (i),(j) 250 nm. For width of 500 nm of there is significant crosstalk, while at 700 nm the image is not well focused, the width of 600 nm is a compromise between these two requirements, notice the slightly different scale on the y axis.

*The problem of reflections and parasitic resonances and tapers*

The abrupt effective-index mismatch between the waveguide and the much wider rectangular MMI section (as in Fig. 5 (a) induces reflections that may interfere with the excitation signal. For a 600 nm wide MMI, we estimate that the oscillation is strongest at the interface and decays to only 1% of the propagating optical power after approximately 15 μm, which is too large from a footprint perspective. This distance can be reduced substantially by using carefully designed adiabatic or parabolic tapers that nearly suppress the reflections. However, even in such an optimized case, the additional footprint remains significant. Here, for simplicity, we adopt straight tapers with a ratio of 1:3. In this case, the oscillation decreases to 1% within only 5μm from the interface, including the taper itself. Tapers with the same 1:3 ratio are also used for the emission waveguides (Fig. 5b), where their role is to funnel the light emitted by the CC into the corresponding waveguides. The wider end of these tapers is chosen to be only 100 nm broader than the waveguide in order to minimize parasitic crosstalk into the emission waveguide.

*1.2 Optimized dimensions of the tapered MMI, the transmission and crosstalk*

The additional tapering modifies the power distribution inside the MMI. The taper acts as a hybrid transition region between the excitation waveguide and the MMI, and therefore also contributes to image formation. As a result, the image positions must be recalculated for the tapered MMI. This can, in principle, be done analytically by dividing the tapered section into a sequence of MMIs with gradually varying widths. Here, however, we use an eigenmode expansion simulator (Lumerical [51] EME), in which the tapered region is modelled using continuously varying cross-sectional sub-cells with a width of 250 nm, as shown in Fig. 5 (b). Fig. 5 (c) and Fig. 5 (d) compare the positions of the first image replica for the rectangular and tapered MMIs. To enable a direct comparison with the rectangular MMI, the image position is referenced to the beginning of the untapered section of the MMI. It can be seen that, in the tapered MMI, the first image is shifted by approximately 130–150 nm towards the input relative to the rectangular case. From this, we infer that roughly half of the taper length acts as an effective input section of the MMI. Based on this result, we estimate the overall dimensions of the tapered MMI to be 1.6 μm × 600 nm for the 150 nm-thick structure and 1.7 μm × 600 nm for the 250 nm-thick structure.

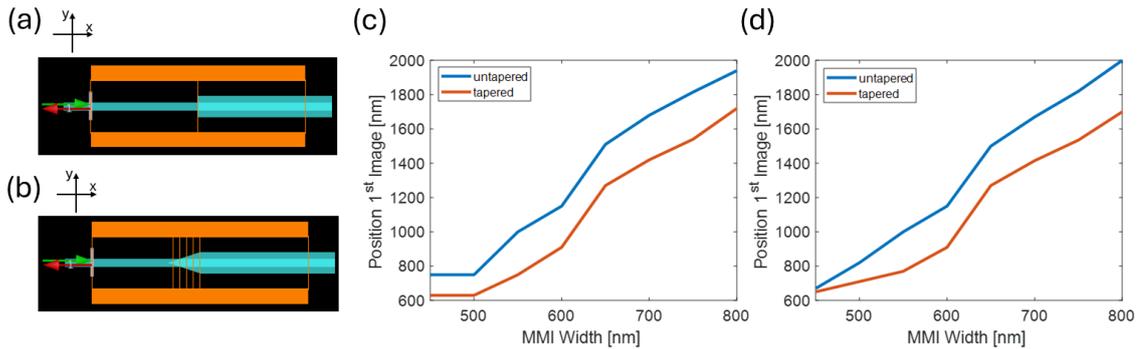

Fig. 5. Schematics for the EME simulations of (a) rectangular and (b) tapered MMIs positions of the first power image replica for straight and tampered MMI with thickness of (c) 150 nm and (d) 250 nm.

In Figs. 6 (a)-(b), we show the power distribution in the central part of a well-calibrated tapered MMI for the two thicknesses. The MMI transmission is calculated by measuring the transmission through the horizontal waveguides to the left and right of the MMI, and is 99% in both cases. Since the image is halfway in the calculation, we can estimate the insertion loss is around 0.5%. Similarly, the crosstalk is calculated by measuring the transmission from the waveguide for excitation to one of the vertical waveguides used to collect emission, and is less than -40 dB.

Finally, we assess the fabrication tolerance of the tapered MMI by introducing small variations in the device length. As shown in Fig. 6 (c), the transmission decreases to 98%, which remains acceptable, for a length variation of ±200 nm, while the crosstalk stays below −30 dB over a band of 600 nm. Using the well known MMI scaling relations [63], $\frac{\delta L}{L} = 2\frac{\delta W}{W} = \frac{\delta \lambda}{\lambda} = \frac{\delta n}{n}$, we estimate that variations of width $\delta W = \pm 40 nm$, $\delta \lambda = \pm 80 nm$ and $\delta n = \pm 0.3$ would lead to performance comparable to that obtained forprovide a similar performance to a variation of length of $\delta L = \pm 200 nm$, thereby defining the tolerance range of the system parameters. We therefore conclude that the width is the most critical MMI dimension, although this tolerance remains readily achievable with modern nanofabrication techniques. The 80 nm operational bandwidth is sufficiently wide to ensure reliable operation of the CC even under environmental perturbations, such as temperature fluctuations or stress. However, it is too narrow to support both resonant and non-resonant excitation of the CC, as the latter occurs at around 530 nm [13].

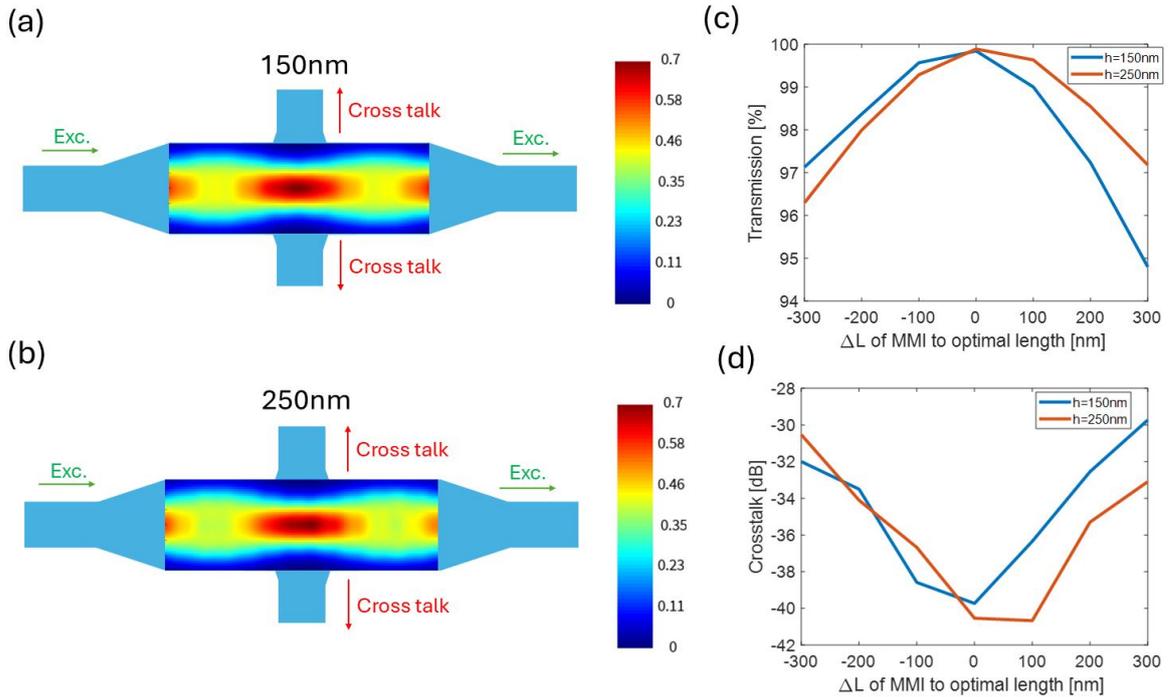

Fig. 6. Size and power distribution for the optimized MMI for (a) 150 nm and (b) 250 nm thickness and relative c) transmission and d) crosstalk for MMI with a ΔL length with respect to the optimized length.

### 1.4 Modelling and emission of the CC

The CC is modelled as an electric dipole whose orientation is set by the crystallographic orientation of the diamond chiplet. The fabrication process is optimized for the <100> direction [36]. For a (100)-oriented diamond chiplet [36], the SnV CC can align along any of the four equivalent <111> axes. A waveguide oriented along ⟨110⟩ provides the most favorable geometry for coupling to the fundamental TE-like mode, since the coupling is governed by the dipole–field overlap at the emitter position. In this configuration, two of the four possible SnV axes lie in the plane transverse to the propagation direction and form an angle of 54.7° with the surface normal.

When the SnV is positioned at the center of the waveguide, these orientations provide the strongest overlap with the local TE field. The excitation rate can be estimated from the local optical intensity and the dipole–field projection at the emitter position. Because the emitter is effectively point-like compared with the diffraction-limited excitation spot, the field can be evaluated at the center of the excitation image.

The excitation of the SnV center is estimated in Lumerical by simulating the pump field at the excitation wavelength and extracting the local electric field at the emitter position with a frequency-domain monitor. The relevant excitation strength is taken as the field component projected onto the SnV dipole axis, $E_\parallel = \hat{\mathbf{u}} \cdot \mathbf{E}$, from which the local intensity is calculated as $I_{\parallel} = \frac{1}{2}nc\varepsilon_0 \mid E_{\parallel} \mid^2$. The absorbed excitation power is then obtained from $P_{\text{abs}} = \sigma_{\text{eff}}I_{\parallel}$, where $\sigma_{\text{eff}} = 10^{-12}$ cm$^2$ is the assumed effective absorption cross section, and the excitation rate is given by $R_{\text{exc}} = P_{\text{abs}}/(h\nu)$. The simulation is normalized to the launched pump power, allowing the excitation to be scaled directly to any input power. In this treatment, the 100 nm$^2$ value is used as an effective absorption cross-section rather than as a physical square area in the numerical model. The cross-section, as already mentioned, is here an assumed effective modelling parameter estimated from literature research, and in in analogy with the absorption cross-section values for NV [64–66] and SiV [67] CCs. Finally, we have optimized the excitation image to be exactly in the center, so we can consider the CC to be excited with the power at the center of the image. The corresponding ZPL emission is then scaled by the quantum efficiency of the SnV CC, $\eta_{QE} = 80\%$ [54], the Debye-Waller factor, $\eta_{DW} = 0.57$ [68], and the branching ratio of the selected ZPL transition, $\eta_{orb} = 0.78$ [69].

We study the emission of CCs located at different positions with respect to the center of the MMI and oriented along ⟨111⟩. The fraction of emitted power propagating into the waveguides is calculated as the ratio between the power recorded by a monitor placed perpendicular to a given waveguide and the total emission of the CC, obtained from the sum of the power recorded by six monitors covering a cube of 50 nm side surrounding the emitter.

For a CC positioned at the center of the cross-shaped MMI, the fraction of emitted power propagating towards each excitation waveguide is 22.5%, while the fraction propagating towards each emission waveguide is about 13%. This difference is likely due to the more favorable tapering of the excitation waveguides, which can channel more light, as well as to internal reflections within the MMI. It is worth noting that, although the dipole emits in all directions, about 71% of the total emitted power is funneled into the four waveguides connected to the MMI, likely as a result of the strong confinement and internal reflections in the MMI region.

In Fig.7 (a)-(c) we show the transmission of the CC towards the emission waveguide in case of misalignment of CC up to ±40 nm in the X, Y or Z direction. The blue curve represents the mere geometrical effect while the red curve is the product of the blue curve with the overlap of the power of the excitation image and the cross section of the CC, shown in Fig. 7 (f), with reference to this figure we can notice that a misalignment in the vertical direction with respect to the center of the power image has different weights depending on the direction, namely 99% for a misalignment of 40 nm in the X direction, 96% in the Y direction, and a 86%, we conclude that the coupling is strongly affected for misalignment in vertical direction.. It can be seen in Fig. 7 (a) the misalignment in the X direction is not particularly relevant, while the misalignment in the Y direction, shown in Fig. 7 (b) creates an asymmetry between the emissions to the corresponding waveguides due to the closer or farter localization of the CC to the output. The vertical misalignment, shown in Fig. 7 (c) is relatively modest geometrically but is strongly dependent on the overlap with the power of the excitation image. Similarly, in In Fig.7 (d), (e), (f) the transmission for the excitation waveguides results overall more stable for horizontal displacements of the CC, varying less than 1% for the considered shifts, while, again the vertical misalignment causes the strongest decrease in transmission because of the poor overlap between the power of the image and cross section of the CC.

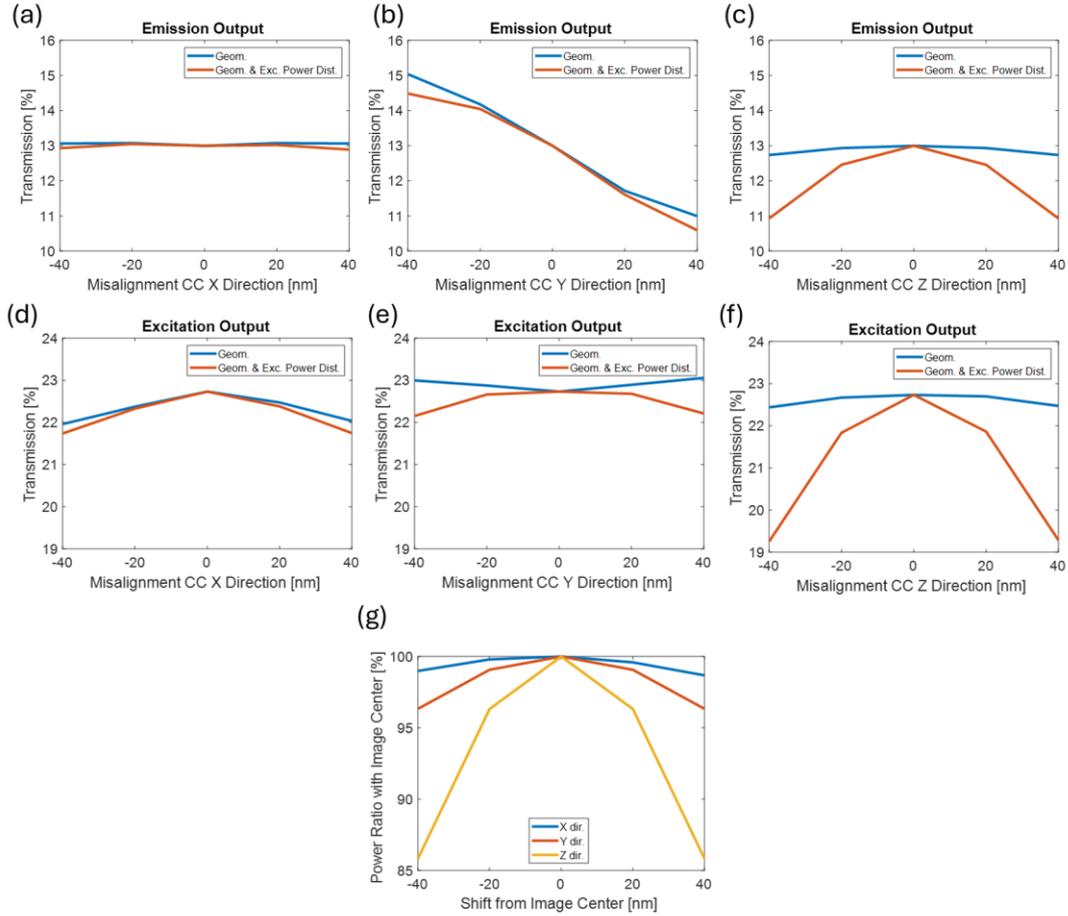

Fig.7. Emission of CC as a function of position supposing 100% coupling, and with coupling normalized by the cross section overlap with the image power for (a) x-misalignment, (b) y-misalignment, (c) z misalignment in the emission waveguide, and (d), (e), (f) in the excitation waveguide, and (g) power ration of the image with respect to the center for the corresponding misalignment.

## 2. Optimization of tethers

Tethers are important because they hold the chiplet to the supporting frame. They must therefore be mechanically robust enough to sustain the chiplet structure, while remaining locally fragile at specific points so that they can be broken by the needle during pick-and-place step. Since they ideally do not carry optical signals, additional features such as alignment facilitators can be integrated into their design, provided sufficient mechanical stability is maintained. Fig. 8 (a) shows a section of the tether highlighting its three main features: the 50 nm connection to the supporting frame, intended to break during pick-and-place, the alignment facilitator; and the intersection with the waveguide. Ideally, light should not propagate from the waveguide into the tethers. In practice, however, the tether–waveguide intersection introduces loss and reflections due to the effective index step. This is similar to the effect discussed for the rectangular MMI, but it is more pronounced here because of the micrometric length of the tether and its larger effective refractive index mismatch with respect to the waveguide. More importantly, the resulting oscillations affect the light transmitted beyond the tether more significantly, as shown in Fig 8 (c), which is particularly relevant because at least one of these paths carries the excitation signal. The tether optimization is therefore focused on maximizing transmission while minimizing oscillations beyond the tether. One possible solution is to use MMI-like structures, as described in the previous paragraph, which can provide high transmission, limited reflections, and strong crosstalk suppression. However, such structures are relatively bulky and may be more susceptible to loss if the optical images are not well focused at the output.

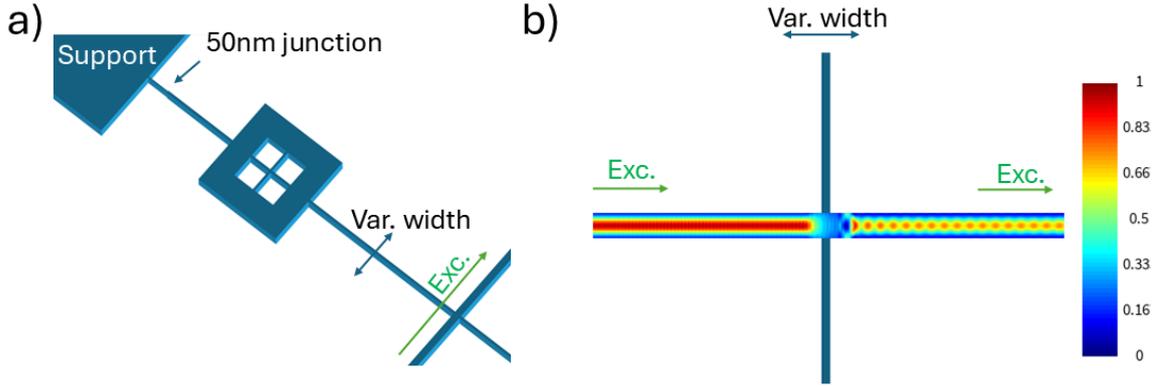

Fig.8. (a) Bird's eye view of the tether with 50 nm junction to the supporting frame and varying width aligned (scale 1:1:2), (b) the oscillation in power in the excitation waveguide after crossing the tether.

In Fig. 9 (a), we show the transmission of the tethers for different tether widths and for thicknesses of 150 nm and 250 nm. As expected, the transmission decreases monotonically with increasing tether width. In Fig. 9 (b) and 9c, we show the oscillation amplitude for different tether widths at distances of 2.5, 5, 7.5, and 10 μm from the tether for both thicknesses. As expected, wider tethers induce stronger oscillations, while the oscillations decrease with increasing distance from the tether. It can be seen that at a distance of 5 μm from the tether, the oscillation amplitude is only slightly larger than at greater distances, and that for 100 nm wide tethers the oscillations remain relatively small (< 2%). This suggests that 100 nm wide tethers, combined with a minimum spacing of 5 μm from other circuit elements, yield sufficiently small oscillations.

The use of tapered tethers [70] could further reduce the oscillations, but at the expense of an undesirably larger footprint. A more practical solution is to place the tethers only on the excitation waveguides. This offers two advantages: first, the signal emitted by the CC remains unaffected by distortions induced by the tethers; second, the excitation power, whether provided by a laser or by other CCs, can be compensated by increasing the input power or the number of single photons, thereby offsetting the tether losses.

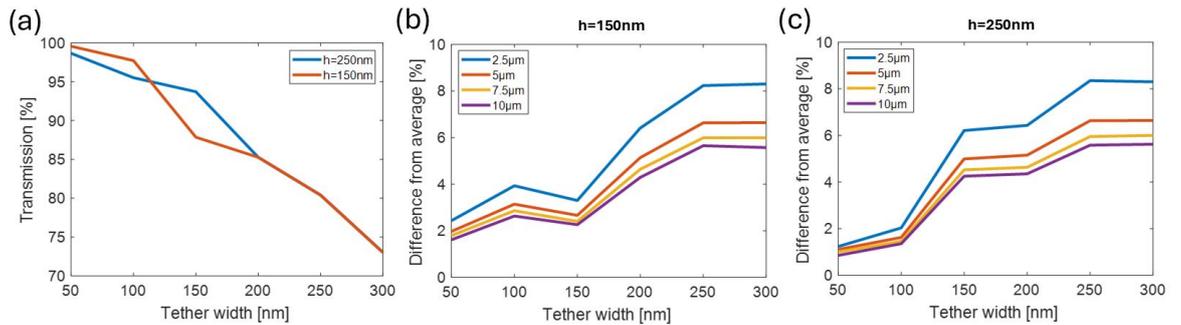

Fig. 9. (a) transmission through tethers of different width, (b) oscillations at 2.5μm, 5μm, 7.5μm and 10μm past the tethers for different width for h= 150 nm and (c) h= 250 nm.

## 3. Optimization of the adiabatic couplers

The coupling between the diamond waveguides of the chiplet and the SiN waveguides of the receptor is the most critical part of the device, as it connects two separate interfaces fabricated using different materials and technologies. To achieve efficient coupling, we employ adiabatic couplers that transfer the optical mode from one waveguide to the other via a gradual change in their geometrical properties, such as width, thickness, or shape, in the coupling region. Here, for the sake of simplicity, we consider a linear reduction of the widths of the overlapping waveguides. In this case, the main parameters to be optimized are the waveguide taper lengths (equivalently, the taper angles) and their overlap shift. As shown schematically in Fig. 10 (a), all tapered waveguides terminate in a 100 nm constriction to ensure nanofabrication reproducibility.

We optimize the adiabatic couplers by varying both the SiN and diamond protrusion lengths to 3 μm, 5 μm, and 7 μm, yielding 9 combinations, and by varying the overlap shift from 0 to 5 μm in steps of 1 μm. This results in relatively long tapers, similar to the structure shown in Fig. 10 (b). For all nine protrusion-length combinations, the overlap shift does not significantly change transmission, indicating that the design is robust against horizontal misalignment. Fig. 11 (a) shows the mean transmission loss, averaged over the overlap shift, for the 150 nm thick waveguide, while Fig. 11 (b) presents the corresponding standard deviation. It can be seen that the combination of 5 μm SiN and 5 μm diamond protrusions provides both the lowest loss, around 1%, and the smallest standard deviation, about 0.01%. To ensure good coupling under both forward and backward misalignment, we optimize the structure for an overlap of 2 μm. Owing to the symmetry of the system, the highest transmission is obtained in the absence of lateral misalignment. The final structure, therefore, consists of 5 μm protrusions on both the SiN and diamond couplers, with a 2 μm overlap. Under ideal conditions, the transmission of this optimized coupler is expected to be about 99%.

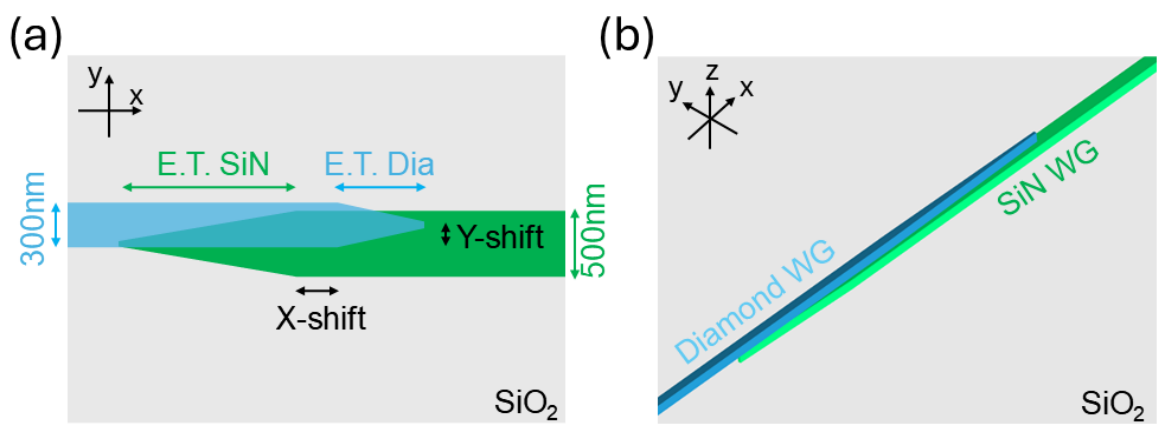

Fig. 10. (a) Adiabatic couplers with the four parameters used for the optimization and (b) bird's eye view of the optimized adiabatic couplers correctly aligned (scale ratio 1:1:2).

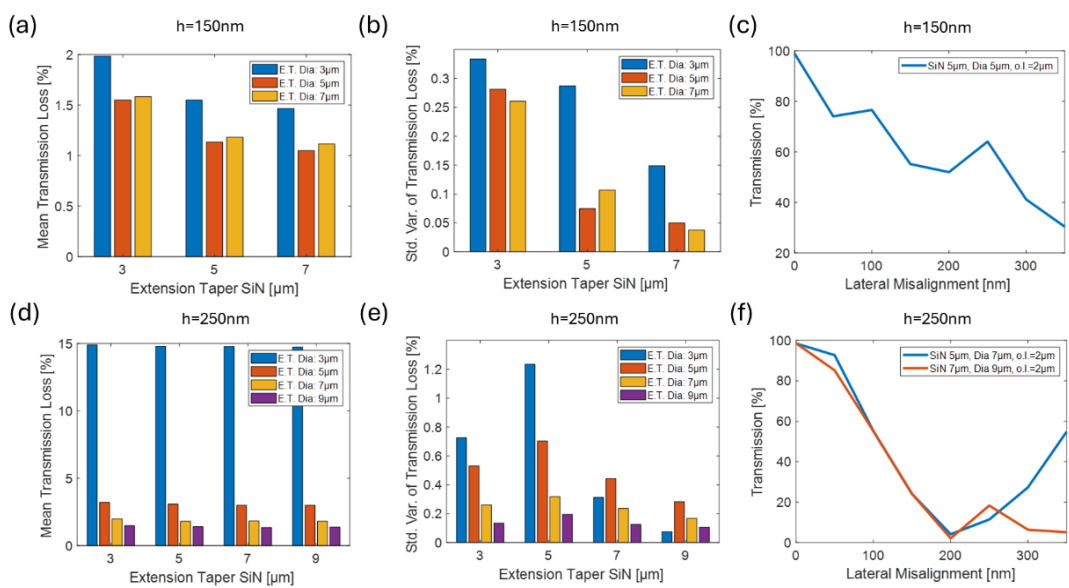

Fig. 11. Process for optimization of adiabatic couplers: (a) mean transmission for various protrusions od SiN and Diamond over their overlap (0µm to 5µm at steps of 1µm), (b) standard deviation of the these transmission over the considered overlap (c) evaluation of side misalignment (off-line shift of couplers) for the optimized adiabatic coupler for h=150 nm and (d), (e), (f) 250 nm.

Lateral off-axis misalignment is the dominant source of loss. As shown in Fig. 11 (c) and (f), a lateral misalignment larger than 150 nm already reduces the transmission to about 50%, and the transmission approaches 20% for a misalignment of 350 nm. To mitigate the loss arising from the inevitable lateral misalignment, several approaches have been reported in the literature, including tapered or more complex taper geometries [71–73], tilted waveguides that trade lateral misalignment for angular misalignment, which is easier to control [74], and the use of intermediate couplers [75]. In the present cross-shaped structure, where the input and output paths are oriented perpendicular to each other, accurate alignment is required in both the $x$ and $y$ directions. This makes approaches such as coplanar coupling ineffective [76–78]. rBy contrast, rotational misalignment is expected to be easier to correct, since the structure is sufficiently large to ensure precise angular alignment between the receptor and the chiplet; we therefore neglect it in this study.

In Figs. 11 (d)–(f), we repeat the analysis for 250 nm thick waveguides. Owing to the different refractive index in this case, we also include a 9 $\mu$m long protrusion, resulting in 16 different combinations. For these waveguides, the standard deviation is larger, and the highest mean transmission is obtained for the combination of a 7 $\mu$m SiN protrusion and a 9 $\mu$m diamond protrusion. However, for compactness, the combination of a 5 $\mu$m SiN protrusion and a 7 $\mu$m diamond protrusion is also attractive, with only about 0.5% lower loss. Also in this case, we choose an overlap of 2 $\mu$m. The lateral misalignment is more critical here, with the transmission approaching zero at 200 nm. Interestingly, for the 5 $\mu$m–7 $\mu$m protrusion combination, we observe a transmission increase, likely due to some hybrid lateral directional coupling between the protrusions. Overall, the directional couplers with 150 nm thick waveguides outperform the 250 nm thick waveguide designs.

## 4. Total transmission

The total transmission is given by the product of the transmission and emission efficiencies of the different components in the circuit,
$$T = T_{dc}T_t T_{wg} Q_{cc} P_{cc} E_{cc} T_{dc}$$

where $T_{dc}$=98-99% is the transmission of each directional coupler, $T_t = 95 - 98\%$ is the transmission through the tether junction, $T_{wg}$ is the transmission through the diamond and SiN waveguides, which is assumed to be close to 100% since the intrinsic waveguide losses are smaller than the other loss contributions. $Q_{cc} = 45.6\%$ accounts for the CC's quantum efficiency and Debye–Waller factor, as reported in the literature. $P_{cc} = 99\%$ is the fraction of excitation power reaching the CC, determined by the overlap between the incoming field and the effective cross section of the CC, while $E_{cc} = 13\%$ is the fraction of emission coupled into one of the emission waveguides.

Under the best conditions, we estimate a total transmission of $T = 5.3$–$5.6\%$, which is in line with the best reported experimental values of about 1% for similar systems [49]. A small vertical misalignment of the CC with respect to the center of the MMI ($\pm 20$ nm) is expected to reduce the transmission by about 0.5%, while a horizontal misalignment between the MMI and the CC would lead to only about 0.1% reduction. The main source of loss, remains the misalignment of the adiabatic couplers, which in the worst case may reduce the transmission to nearly zero. Particular attention must therefore be paid to optimizing the chiplet's placement on the receptor.

## 5. Optimization of placing of chiplet on receptor

### 5.1 The mechanical properties of tethers to sustain complex structures

As mentioned above, the tethers serve two purposes. First, they must support the structure, which requires sufficient stiffness and appropriate dimensions to keep the suspended device flat and to limit bending or buckling, thereby preserving planarity. Second, they must break at predefined locations during the pick-up step; this is achieved by introducing a local constriction at the intended breaking point.

At the same time, this constriction must still be able to sustain the weight of the structure and remain mechanically stable. To verify that the suspended device does not undergo significant flexural or torsional deformation, we consider a conservative worst-case scenario in which a single tether supports the full weight of the structure, applied as a point load $F$ at the center of the tether, i.e. at its most compliant point. This configuration is more accurately described as a doubly clamped beam with a central point load, which can be treated analytically using Timoshenko beam theory. The second moment of inertia of the cantilever is given by $I_{xx} = wh^3/12$, where $w = 50\ nm$ is the width of the constriction and $h = 150\ nm$ or $250\ nm$ is its thickness. Using a Young modulus of 1050GPa [79] and a density of ρ=3.52g/cm³ or 3.52*10⁻¹⁵ Kg/μm³ [80], and estimating the total weight of the device from the volume of its components, we obtain a force of $F = 4 \times 10^{-8}$N in the center of the cantilever. The resulting central deflection, given by $\delta = (Fa^3b^3)/(3EI_{xx}L^3)$, is only about 2nm. The corresponding stress, assuming that the force acts over an area of $50\ nm \times 50\ nm$, remains well below the fracture limit (100 MPa), indicating that the structure can safely withstand the applied load [81–83].

*5.2 The use of structures to facilitate alignment during structure placement*

As discussed in Section 3, accurate alignment of the chiplet on the receptor is crucial to ensure efficient light transmission into the chiplet and back through the adiabatic couplers. Pick-and-place technology, however, is inherently prone to misalignment, since the operation is often performed at least partly by a human operator. To improve the alignment accuracy, we introduce alignment-facilitating structures on the chiplet tethers, as already shown in Fig. 8 (a), together with their complementary structures on the receptors, as shown in Fig. 12 (a). These structures consist of a 4 μm × 4 μm square with a 2 μm × 2 μm opening in the center. The 1 μm-wide sidewalls of the hollow square are readily visible under a modern optical microscope, allowing alignment by superimposing the edges of identically sized openings on the chiplet and the receptor. The central opening also allows direct visual inspection during positioning, which is particularly useful given that the 100 nm waveguides cannot be resolved with standard optical microscopy. Figs 12 (b) and 12 (c) show the facilitators from the top view and from a bird's-eye view when properly aligned.

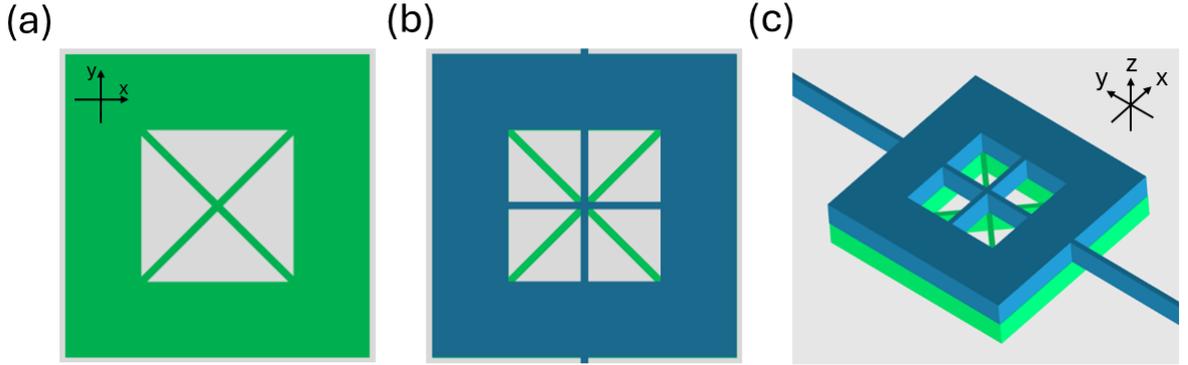

Fig. 12. (a) The receiver alignment facilitator structure on SiO2 and (b) with the diamond facilitator structure correctly placed hereon, as seen from a top view and (c) bird's eye view (scale ratio 1:1:2).

Finally, Fig. 13 shows the system after correct placement of the diamond chiplet on the SiN receiver. The chiplet is stably supported by the four alignment facilitators and the four adiabatic couplers, resulting in a mechanically robust structure that is well attached to the receiver. For the 250 nm-thick waveguide, the overall footprint dimensions are $L_x = 49.9\ \mu m$ and $L_y = 38.6\ \mu m$. Here, $L_x$ is given by the sum of the horizontal contributions from the directional coupler, overlap region, tether section, tether-to-MMI spacing, and half of the MMI length, while $L_y$ is determined by the corresponding vertical contributions, including the SiN directional coupler, the coupler-to-MMI spacing, and half of the MMI width. The parameters used are $L_{dc,dia} = 5\ \mu m$, $L_{dc,SiN} = 7\ \mu m$, $L_{dc,ov} = 2\ \mu m$, $L_{dc,tether} = 5\ \mu m$, $L_{tether} = 0.1\ \mu m$, $L_{tether-MMI} = 5\ \mu m$, $L_{MMI/2} = 0.85\ \mu m$, $L_{dc-MMI} = 5\ \mu m$, and $W_{MMI/2} = 0.3\ \mu m$.

For the 150 nm-thick waveguide, the same expressions apply, but with $L_{dc,SiN} = 5\ \mu m$ and $L_{MMI/2} = 0.8\ \mu m$, resulting in $L_x = 45.8\ \mu m$ and $L_y = 34.6\ \mu m$. Therefore, the footprint is around $45 \times 35$ μm². It is important to note that the SiN adiabatic couplers are positioned underneath the chiplet, while the tethers and diamond facilitators remain within the chiplet area defined by the corresponding excitation and emission waveguides. Therefore, these dimensions represent the footprint of the complete structure.

The structure can also be readily integrated with other optical devices. The excitation light traveling to and beyond the CC can still be used for other purposes, such as exciting additional CCs in similar structures, while the two emission waveguides provide two independent outputs for the signal emitted by the CC, which is attractive for quantum communication and quantum internet applications.
.

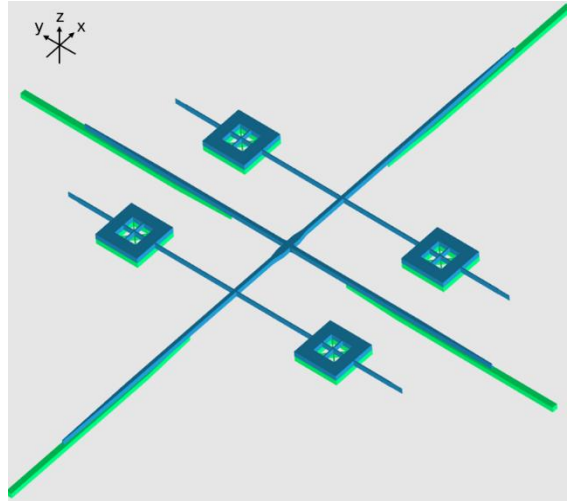

Fig.13. Bird's eye view of the chiplet correctly placed on the SiN receiver, with alignment facilitator structures and markers perfectly matching. (scale ratio 1:1:2).

## 6. Conclusion

In this study, we proposed a simple methodology for optimizing a cross-waveguide diamond chiplet containing a CC, together with the corresponding SiN receiver for pick-and-place integration, guided by device performance, ease of fabrication, footprint, and reduction of available degrees of freedom. For each component, we discussed its role, the associated challenges and limitations, provided a brief overview of the relevant options in the literature, and justified the strategic choices adopted here. More specifically, we selected 150 nm and 250 nm thicknesses for 300 nm-wide diamond waveguides to support one and two TE modes, respectively. We then optimized the tapered multimode interferometer at the center of the cross to achieve focused excitation of a CC positioned at its center, with an excitation efficiency above 99%, crosstalk below −40 dB, reflections below 1%, and good fabrication tolerance over the width and length variations of $\delta W = \pm 40\ nm$ and $\delta L = \pm 200\ nm$, respectively. The fraction of CC emission coupled into each emission waveguide is about 13% and remains sufficiently tolerant to CC misalignment of the order of $\delta z = \pm 20\ nm$, consistent with modern implantation technology. We further analyzed the supporting tethers and found that a 100 nm-wide tether introduces only a 2% oscillation at 5 μm beyond the tether and maintains a transmission above 95%. Finally, we optimized the adiabatic couplers for both the diamond chiplet and the receiver, achieving a transmission above 98% and very good horizontal tolerance. We identified lateral misalignment of the adiabatic couplers as the most critical source of loss and therefore introduced alignment facilitators on both the tethers and the receiver to support accurate placement. Overall, we designed a cross-waveguide device with a compact footprint (≤ 50 μm × 39 μm), an expected nominal CC emission above 5%, and good fabrication and parameter tolerance. The 150 nm-thick device offers a smaller footprint (~ 45 μm × 35 μm) and better coupling to the receiver, whereas the 250 nm-thick design provides lower crosstalk and less lossy structures. The cross-waveguide geometry enables separation of the excitation signal from the CC emission. Since the proposed design is suitable for both uniform DOI substrates and bulk diamond, which are widely used platforms for color-center integration, we expect that this methodology will be of interest to the community working on diamond CCs for quantum communication and computation, as well as on heterogeneous integration schemes.

**Funding.** joint research program "Modular quantum computers" by Fujitsu Limited and Delft University of Technology, co-funded by the Netherlands Enterprise Agency under project number PPS2007.

**Acknowledgements** Dr. Alessio Miranda would like to thank Dr. Erwin van Zwet, Mrs. Elena Volkova and Mr. Christian Primavera for insightful discussions.

We gratefully acknowledge support from the joint research program "Modular quantum computers" by Fujitsu Limited and Delft University of Technology, co-funded by the Netherlands Enterprise Agency under project number PPS2007.

**Disclosures** "The authors declare no conflicts of interest."

**Data availability.** Data underlying the results presented in this paper are not publicly available at this time but may be obtained from the authors upon reasonable request.